\documentstyle[12pt,epsfig]{article}
\date{\today}

\title{  Spin Correlations in top quark pair production
near threshold at the $ e^- e^+ $ Linear Collider }
\author{ Qing Jun Zhang, Chong Sheng Li\footnote{csli@pku.edu.cn}, Jian Jun Liu, and Li Gang Jin\\
{\small Department of Physics, Peking University, Beijing 100871,
China} \\
}
\date{\today}

\begin{document}
\maketitle
\begin{abstract}
  We investigate the spin correlations in top quark pair production near threshold  at the
   $ e^- e^+ $ linear collider.  Comparing with the results above the threshold region,
    we find that near the threshold region the off-diagonal basis, the optimized decomposition of the top quark spins
    above the threshold region, does not exist,  and  the beamline basis is the optimal basis,
    in which there are the dominant spin components: the
up-down (UD) component for $e_L^- e^+$ scattering and the down-up
(DU) component for $e_R^- e^+$ scattering can make up more than
50$\%$ of the total cross section, respectively.

\end{abstract}

\vspace{1.5cm} \noindent  Keywords: top quark, spin correlations,
threshold.


\newcommand{\beq}{\begin{eqnarray}}
\newcommand{\eeq}{\end{eqnarray}}

\newcommand{\ssa}{\sin \theta_W}
\newcommand{\ssb}{\sin^2 \theta_W}
\newcommand{\cca}{\cos \theta_W}
\newcommand{\ccb}{\cos^2 \theta_W}

\newcommand{\tup}{t_{\uparrow}}
\newcommand{\td}{t_{\downarrow}}
\newcommand{\tbup}{\bar{t}_{\uparrow}}
\newcommand{\tbd}{\bar{t}_{\downarrow}}
\newcommand{\uu}{t_{\uparrow}\bar{t}_{\uparrow}}
\newcommand{\dd}{t_{\downarrow}\bar{t}_{\downarrow}}
\newcommand{\ud}{t_{\uparrow}\bar{t}_{\downarrow}}
\newcommand{\du}{t_{\downarrow}\bar{t}_{\uparrow}}

\newpage
The top quark is the heaviest known particle in the standard model
(SM) and rapidly decays before it hadronizes \cite{dec}, and the
spin information of the top quark is preserved from production to
decay. Thus we can expect the spin orientation of the top quark to
be observable experimentally. The spin correlations for the top
quark pair production above the threshold at $e^+ e^-$ colliders
have been extensively discussed \cite{tt2}. Parke and Shadmi
\cite{imp3} proposed the generic spin basis and found that the
``off-diagonal" basis, a special case of the generic spin basis,
is a more optimized decomposition of the top quark spins for $e^+
e^-$ colliders. As shown in Refs. \cite{spi1,spi}, the
off-diagonal basis is indeed the optimal spin basis even after the
inclusion of ${\cal O}(\alpha_s)$ QCD corrections:  at
$\sqrt{s}=400$ GeV using the off-diagonal basis the dominant spin
components in both $e_L^-$ and $e_R^+$ scattering make up more
than $99\%$ of the total cross section at both tree and one-loop
level, but such fraction is only $\sim 53\%$ in the helicity
basis.

Up to now, all studies of spin correlations of the top quark pair
production at the $ e^- e^+ $ collider are limited only to the
process of  $t \bar t $ production above the threshold region. In
the future $e^- e^+ $ linear collider, threshold production of the
top quark  pair will allow to study their properties  with
extremely high precision. Because of large top quark mass and
decay width, the bound-state resonances lose their separate
identify and smear together into a broad threshold enhancement
\cite{thr1}, and as a result, the nonpretubative QCD effects
induced by the gluon condensate are small, allowing us to
calculate the cross section with high accuracy by using
perturbative QCD even in the threshold region. And it is
interesting to investigate the spin correlations of the top quark
pair production near threshold. Some methods used to deal with the
behavior of top quark pair production near threshold have been
established. The Green function technique was demonstrated
suitable to calculate the total cross section and to predict the
top quark momentum distribution, and independent approaches were
developed for solving the Schr\"{o}dinger equation in position
space and the Lippmann-Schwinger equation in momentum space
\cite{thr1,imp2,thr,thr2}.

In this letter, we study  the spin correlations  in the top quark
pair production near threshold. In order to include effects of the
quark-antiquark potential and the decay width in our calculation,
we use two ingredients: the first is the vertex function, which
represents the QCD binding effects and the anomalous interactions,
and can be obtained through solving Lippmann-Schwinger equation
\cite{edms}; the second is the two unstable-particle phase space
 in the nonrelativistic limit \cite{imp2}, which,  combined with the
vertex function,  describes the top  quark momentum distribution.
As a cross check of our calculation, after using above
ingredients,  also we obtain the total cross section and the
momentum distribution which are consistent with ones given in Ref.
\cite{thr}.

In the SM, we consider the process \beq e^- e^+ \to t \bar t
\label{eett} \eeq at the LC with $\sqrt{s}\simeq 2 m_t$. The tree
level $Ve^-e^+$ vertex can be written as \beq \Gamma^\mu_{Vee}
=\gamma^\mu (K_L^V P_- + K_R^V P_+) \label{vee}, \eeq where
$P_{\pm}=(1\pm \gamma_5)/2$, and the SM values for these coupling
factors are $K_{L,R}^\gamma=-e$ for $V=\gamma$, $K_L^Z=e
(2\ssb-1)/2\ssa \cca$ and  $K_R^Z =e \ssa/\cca$ for $V=Z$, and the
$\theta_W$ is the Weinberg angle.

The $Vt\bar t$ vertex function  $\Gamma_{Vt \bar t}$ ($V=\gamma,
Z$) can be generally  written as \beq \Gamma_{V t\bar{t}}^{\mu}& =
& \gamma^\mu(A_V P_{-} + B_V P_+) + \gamma^5 \frac{p^\mu }{m_t}C_V
, \label{vttb} \eeq where $p^\mu$ is the momentum of the outgoing
top quark, and $A_V, B_V, C_V$ are the form factors. As shown in
Fig.1, the vertex function $\Gamma_V$ ($\Gamma^{\mu}_{Vt \bar t}
$) satisfies following integral equation  \cite{imp2}:\beq
\Gamma_V = X_V + \int \frac{d^4k}{(2 \pi)^4} (-4 \pi C_F \alpha_s)
D_{\mu \nu}(p-k) \gamma^\mu S_F(k+\frac{q}{2})
\Gamma_V(k,q)S_F(k-\frac{q}{2}) \gamma ^\nu, \label{aaaa}\eeq
where $X_\gamma=\gamma_\mu$ and $X_Z = \gamma_\mu \gamma_5$, $S_F$
is the top quark propagator, and $D_{\mu \nu}$ is the gluon
propagator. In the nonrelativistic limit, the propagators are
replaced by \beq -4 \pi C_F \alpha_s D_{\mu \nu}(p)
&\longrightarrow& i V(\vec{p}) \delta_{\mu 0} \delta_{\nu 0} , \\
S_F(k+\frac{q}{2}) &\longrightarrow& i
\frac{\frac{1+\gamma^0}{2}-\frac{\vec{k} \cdot \vec{\gamma}}{2
m_t} } {\frac{E_{NR}}{2}+k_0-\frac{\vec{k}^2}{2 m_t} +i
\frac{\Gamma_t }{2}} , \\ S_F(k-\frac{q}{2}) &\longrightarrow& i
\frac{\frac{1-\gamma^0}{2}-\frac{\vec{k} \cdot \vec{\gamma}}{2
m_t} } {\frac{E_{NR}}{2}-k_0-\frac{\vec{k}^2}{2 m_t} +i
\frac{\Gamma_t }{2}} ,\eeq where  $E_{NR}=\sqrt{s}-2m_t$. With
above approximations, when the QCD binding effects and the
CP-violating anomalous couplings are included \cite{edms}, the
form factors $(A_V, B_V, C_V)$ in the vertex function $\Gamma_V$
are given by \beq A_\gamma
& = & \frac{2}{3}e (1- \frac{2C_F \alpha_s}{\pi}) G(|\vec{p_t}|) \varphi  \label{abca} ,\\
A_Z & = & \frac{e}{ \ssa\cca}((\frac{1}{4}-\frac{2}{3}\ssb)(1-
\frac{2C_F \alpha_s}{\pi} ) G(|\vec{p_t}|) \varphi \nonumber \\ &
& +\frac{1}{4}(1-\frac{C_F
\alpha_s }{\pi})F(|\vec{p_t}|) \varphi) , \\
B_\gamma & = &\frac{2}{3}e (1- \frac{2C_F \alpha_s}{\pi})
G(|\vec{p_t}|)  \varphi ,
\\B_Z & = & \frac{e}{4\ssa\cca}((1-\frac{8}{3}\ssb)(1- \frac{2C_F
\alpha_s}{\pi}) G(|\vec{p_t}|)  \varphi \nonumber \\
& &-(1-\frac{C_F
\alpha_s }{\pi})F(|\vec{p_t}|) \varphi) , \\
C_\gamma & = &-ied_{t\gamma}(1-\frac{C_F \alpha_s
}{\pi})F(|\vec{p_t}|)\varphi+\frac{2}{3}ied_{tg}D(|\vec{p_t}|)
\varphi , \\
C_Z & = & \frac{e}{ \ssa\cca}(-id_{tz}(1-\frac{C_F \alpha_s
}{\pi})F(|\vec{p_t}|)\varphi+id_{tg}(\frac{1}{4}-\frac{2}{3}\ssb)D(|\vec{p_t}|)
\varphi ) , \label{abcd} \eeq  where  $d_{t \gamma}$, $d_{tz} $
and $d_{tg}$ are the anomalous couplings of the top quark, and
$\varphi = \frac{|\vec{p_t}|^2}{m_t}-(E_{NR}+i\Gamma_t)$. The
Green-functions (G, F and D) in Eqs.(\ref{abca})-(\ref{abcd}) are
the solutions of the Lippmann-Schwinger equations \cite{edms}:
\beq &&
(\frac{|\vec{p}|^2}{m_t}-(E_{NR}+i\Gamma_t))G(E_{NR},|\vec{p}|)
  + \int \frac{d^3k}{(2\pi)^3}[V(|\vec{p}-\vec{k}|)G(E_{NR},|\vec{k}|)]=1\label{gr}, \\
 && (\frac{|\vec{p}|^2}{m_t} - (E_{NR}+i\Gamma_t))\vec{p^i} F(E_{NR},|\vec{p}|)
  + \int \frac{d^3 k}{(2\pi)^3}[ V(|\vec{p}-\vec{k}|)\vec{k}^i F(E_{NR},|\vec{k}|) ]
   = \vec{p^i}\label{fr}, \\ && (\frac{|\vec{k}|^2}{m_t} - (E_{NR}+i\Gamma_t) ) \vec{p^i}D(E_{NR},|\vec{p}|)
  + \int \frac{d^3 k}{(2\pi)^3}
    [ V(|\vec{p}-\vec{k}|)\vec{k}^i D(E_{NR},|\vec{k}|) ]
    \qquad\qquad \nonumber \\  && \hspace{6.0cm} = \int \frac{d^3 k}{(2\pi)^3}
    [ V(|\vec{p}-\vec{k}|) (\vec{p}-\vec{k})^i G(E_{NR},|\vec{k}|) ]\label{dr},\eeq
  which  can be derived from Eq.(\ref{aaaa}), and the QCD potential
  $V(|\vec{p}-\vec{k}|)$ is  given by \cite{pot} \beq V(|\vec{p}-\vec{k}|) =
 V_C(|\vec{p}-\vec{k}|) + V_{BF}(|\vec{p}-\vec{k}|) + V_{NA}(|\vec{p}-\vec{k}|)\eeq
 with
 \beq   V_C(|\vec{p}-\vec{k}|)&=&-\frac{4 \pi C_F
 \alpha_s(|\vec{p}-\vec{k}|^2)}{|\vec{p}-\vec{k}|^2} [1+(\frac{\alpha_s(|\vec{p}-\vec{k}|^2)}{4 \pi}) a_1 +
 (\frac{\alpha_s(|\vec{p}-\vec{k}|^2)}{4 \pi})^2 a_2 ], \\
V_{BF}(|\vec{p}-\vec{k}|)&=&-4 \pi C_F
 \alpha_s(|\vec{p}-\vec{k}|^2)[\frac{(\vec{p} \times \vec{k})^2 }{m^2_t
 |\vec{p}-\vec{k}|^4} \nonumber +\frac{1}{4 m^2_t}
 +\frac{3 i (\vec{p}\times \vec{k}) \cdot (\vec{S_t}+\vec{S_{\bar t}})}{2 m^2_t
 |\vec{p}-\vec{k}|^2}\\
 &&+\frac{1}{2m_t^2}((\vec{S_t}+\vec{S_{\bar t}})^2-\frac{((\vec{S_t}+\vec{S_{\bar t}})
  \cdot (\vec{p}-\vec{k}))^2}{|\vec{p}-\vec{k}|^2})
 ], \\
V_{NA}(|\vec{p}-\vec{k}|)&=& -\frac{3 \pi^2 C_F
\alpha_s^2(|\vec{p}-\vec{k}|^2)}{m_t |\vec{p}-\vec{k}| },\\
\alpha_s(|\vec{p}-\vec{k}|^2)& = & \frac{4 \pi}{\beta_0
\ln(|\vec{p}-\vec{k}|^2/\Lambda^2)}[1-\frac{2
\beta_1}{\beta_0^2}\frac{\ln(\ln(|\vec{p}-\vec{k}|^2/\Lambda^2))}{\ln(|\vec{p}-\vec{k}|^2/\Lambda^2)}],
 \eeq where $\vec S_{t}$ and $\vec S_{\bar t}$ are the top and antitop spin operators, and
  \beq  a_1&=&43/9, \ \ a_2 = 155.842, \ \ \Lambda = 226 \; {\rm MeV},\nonumber \\
\beta_0 &=& 23/3, \ \
 \beta_1=58/3, \ \ \mu= 20 \;  {\rm GeV}. \eeq

Using above vertex functions, we calculate the spin correlations.
In the $t\bar t$ center of mass frame (CMS), the scattering plane
is defined to be the X-Z plane where the electron is moving along
the $+Z$ direction and $\theta_t$ is defined as  the scattering
angle of the top quark, and we also set $\phi_t=0$. The Born
helicity amplitudes for the process (\ref{eett}) are obtained by
summing the contributions from both the $Z$ and $\gamma$: \beq
M(h_{e^-}, h_{e^+}, h_{t}, h_{\bar{t}})=
 M(h_{e^-}, h_{e^+}, h_{t},
h_{\bar{t}})^\gamma + M(h_{e^-}, h_{e^+}, h_{t}, h_{\bar{t}})^Z
R(s), \label{aht} \eeq where $s=4E_e^2$ is the total energy in
CMS, $E_e$ is the energy of the electron,  and $R(s)= s/(s-M_Z^2)$
\footnote{At NLC with $\sqrt{s}\simeq 2m_t$, the imaginary part of
the $Z$ propagator can be neglected safely.}.

In the  generic spin basis \cite{imp3} the top quark (anti-top
quark) spin states are defined in the top quark (anti-top quark)
rest-frame, where one decomposes the top (anti-top) spin along the
direction $\hat{s}_t$ ($\hat{s}_{\bar t}$), which makes an angle
$\xi$ with the anti-top (top) momentum in the clockwise direction.
Thus, the state $\tup \tbup$ ($\td \tbd$) refers to a top with
spin in the $+\hat{s}_t$ ($-\hat{s}_t$) direction in the top
rest-frame, and an anti-top with spin $+\hat{s}_{\bar t}$
($-\hat{s}_{\bar t}$ ) in the anti-top rest-frame.

In the generic spin basis, the amplitudes $M(h_{e^-}, h_{e^+},
\hat{s_{t}}, \hat{s_{\bar{t}}})$ for the process $e^- e^+ \to t
\bar t$ can be generally written as \beq M(-+\uu\; {\rm or} \;
\dd)& = & \pm 2 E_e  [ m_t (A_L+B_L) \sin{\theta}\cos{\xi}- (|\vec
{p_t}| (A_L-B_L) \nonumber \\ && + \cos{\theta} E_e
(A_L+B_L))\sin{\xi}\pm 2 E_e C_L \frac{|\vec {p_t}| }{m_t}
\sin{\theta} ],\label{mg1}
\\
M(-+\ud\:{\rm or} \; \du)&=& 2 E_e [ m_t (A_L+B_L) \sin{\theta}
\sin{\xi} \pm (E_e (A_L+B_L) + |\vec
{p_t}|(A_L-B_L)\cos{\theta} ) \nonumber \\
&& + \cos{\xi} (|\vec {p_t}|(A_L-B_L)+ E_e (A_L+B_L) \cos{\theta})
],\label{mg2}
\\
M(+-\uu\; {\rm or} \; \dd)& =& \pm 2 E_e  [ m_t (A_R+B_R)
\sin{\theta} \cos{\xi} \nonumber \\ && + ((A_R-B_R) |\vec {p_t}| -
\cos{\theta} E_e (A_R+B_R))\sin{\xi}\mp 2 E_e C_R \frac{|\vec
{p_t}|}{m_t} \sin{\theta} ],\label{mg3}
\\
M(+-\ud\:{\rm or} \; \du)& = & 2 E_e [m_t (A_R+B_R) \sin{\theta}
\sin{\xi} \mp (E_e (A_R+B_R) + |\vec
{p_t}| (B_R-A_R) \cos{\theta}) \nonumber \\
&& +(|\vec {p_t}|(B_R-A_R) +E_e (A_R+B_R) \cos{\theta}) \cos{\xi}
] ,\label{mg4} \eeq where  $A_{L,R}$, $B_{L,R}$,  $C_{L,R}$ are
the form factors and defined as \beq A_{L,R}& = &\frac{1}{s}\left
(K^{\gamma}_{L,R}A_\gamma
+ K^{Z}_{L,R}A_Z R(s)\right ), \label{alr}\\
B_{L,R}& =&\frac{1}{s}\left ( K^{\gamma}_{L,R}B_\gamma +
K^{Z}_{L,R}B_Z R(s) \right ),\label{blr}
\\ C_{L,R}&=&\frac{1}{s}\left ( K^{\gamma}_{L,R} C_\gamma + K^{Z}_{L,R} C_Z R(s) \right ). \eeq
Because the $V t \bar t$ vertex (V=$\gamma$,Z) has complex
structures near threshold, we can not find such a spin angle $\xi$
that makes Eq.(24) equal to zero, and then there is not the
off-diagonal basis, in contrast to the case of the above
threshold. The amplitudes in  the helicity basis and the beamline
basis can be obtained
 by setting $\cos \xi =\pm 1$ and  $\cos \xi = (\cos \theta +
\frac{|\vec {p_t}|}{m_t})/(1+ \frac{|\vec {p_t}|}{m_t}\cos \theta
)$ in Eqs.(\ref{mg1})-(\ref{mg4}), respectively.

In the nonrelativistic limit, the differential cross sections of
two unstable particle production \cite{imp2}  in the generic spin
basis can be expressed as \beq \frac{d\sigma(h_{e^-}, h_{e^+},
\hat{s_{t}}, \hat{s_{\bar{t}}})}{d\cos{\theta}} &=&
\frac{\Gamma_t}{8 \pi^2 M_t^2} \int\frac{|\vec
{p_t}|^2}{(E_{NR}-\frac{|\vec {p_t}|^2}{m_t})^2+\Gamma_t^2}|
M(h_{e^-}, h_{e^+}, \hat{s_{t}}, \hat{s_{\bar{t}}})|^2d|\vec
{p_t}|\label{dcsa}. \eeq

 In the numerical calculation, we use the following parameters
as standard input \cite{par}: \beq \alpha_s(M_Z) &=&0.117, \ \
m_Z=91.188 \; {\rm GeV}, \ \ m_t=174 \; {\rm GeV}, \nonumber \\
\Gamma_t &=& 1.43 \; {\rm GeV}, \ \ \sin^2{\theta_W}=0.2311. \eeq

We define the fractions of the total cross sections for the
different spin components in the beamline spin basis as following:
\beq R(e^-_{L,R}, e^+, \hat{s_{t}}, \hat{s_{\bar{t}}}) =
\frac{\sigma(e^-_{L,R}, e^+, \hat{s_{t}},
\hat{s_{\bar{t}}})}{\sigma_{total} (e^-_{L,R} e^+ \to t \bar t) }.
\eeq With $E_{NR}=5$ GeV, we have \beq R(e_{L}^- e^+ \to \ud \;
{\rm or} \; e_{R}^- e^+ \to \du ) & \simeq & 50\%,
\\ R(e_{L}^- e^+ \to \du \; {\rm or} \; e_{R}^- e^+ \to \ud) & \simeq & 24\%,
\\ R(e_{L}^- e^+ \to \uu \; {\rm or} \; e_{R}^- e^+ \to \dd) & \simeq & 13\%, \\
R(e_{L}^- e^+ \to \dd \; {\rm or} \; e_{R}^- e^+ \to \uu) & \simeq
& 13\%. \eeq These results show that in the $t \bar t$ threshold
region all spin components can not be neglected. In Fig.2 we show
the differential cross sections for the precess $e^-_{L,R} e^+ \to
t \bar t$ in the beamline basis with $E_{NR} = 5$ GeV. One can see
that there is a dominant spin component when scattering angle
$\theta$ ranges between $\pi /3$ and $ 2 \pi /3 $. More precisely,
according to the definition of R in Eq.(33), we integreted the
$\theta$  from $\pi /3$ to $ 2 \pi /3$, instead of 0 to $ \pi $,
and then have $ R(e_{L}^- e^+ \to \ud \; {\rm or} \; e_{R}^- e^+
\to \du) \simeq 79 \%$. But, as shown in Fig.3, in the helicity
basis there are not such dominant spin components.

Moreover, in our calculation, we considered the Higgs potential
effect on the vertex functions. Our numerical results show that
such effect is very small and can be neglected. The anomalous
couplings ($d_{t \gamma}$, $d_{tz}$, $d_{tg}$) are one of several
sources to provide CP-violation \cite{cpv}, and the numerical
results show that these effects on the spin correlations in the $t
\bar t $ threshold region are very small, too.  For example, with
taking $d_{t \gamma}$= $d_{tZ}$ = $d_{tg}$= $10^{-3}$ \cite{edms},
the corresponding changes of the spin correlations are smaller
than 0.1$\%$.

To summarize, we have calculated the spin correlations in the top
quark pair production near threshold at the $ e^- e^+ $ Linear
Collider in the SM. We start from the general form of the $V t
\bar t$ vertex (V=$\gamma$,Z) near threshold, derive out the
amplitudes in the generic spin basis, and give the differential
cross sections in the NNLO QCD potential. Comparing with the
previous results above the threshold region in Refs.
\cite{spi1,spi}, we find:

\begin{quotation}

(a) The most important difference between the two regions is that
in the above  threshold region we can find the off-diagonal basis
in which only one spin component is appreciably non-zero, but in
the threshold region the off-diagonal basis does not exist.

(b)  Near threshold  the beamline basis is the optimal basis, in
which there are the dominant spin components: the up-down (UD)
component for $e_L^- e^+$ scattering and the down-up (DU)
component for $e_R^- e^+$ scattering can make up more than 50$\%$
of the total cross section, respectively.

(c)  The observables of the spin correlations near threshold are
less advantageous than ones of above the threshold region.
Nevertheless, because of the extremely high measurement precision
of the top quark pair threshold production, it is still valuable
to study the spin correlations in the top quark pair production
near threshold.
\end{quotation}

\vspace{2.5cm}

\hspace*{4.5cm}ACKNOWLEDGMENTS\\

This work was supported in part by the National Natural Science
Foundation of China.

\vspace{1cm}

\newpage
   \begin{figure}[t]
  \hspace*{\fill}
\centerline{\epsfig{file=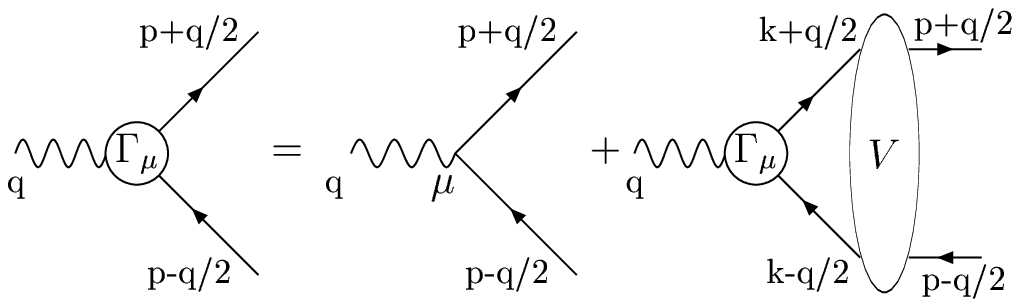, width=400pt}}
\caption{\small
      Lippmann-Schwinger equation in diagrammatical form.
      \label{fig:ladder}
  }
  \hspace*{\fill}
\end{figure}

\newpage
\begin{figure} \label{pa}
\centerline{\epsfig{file=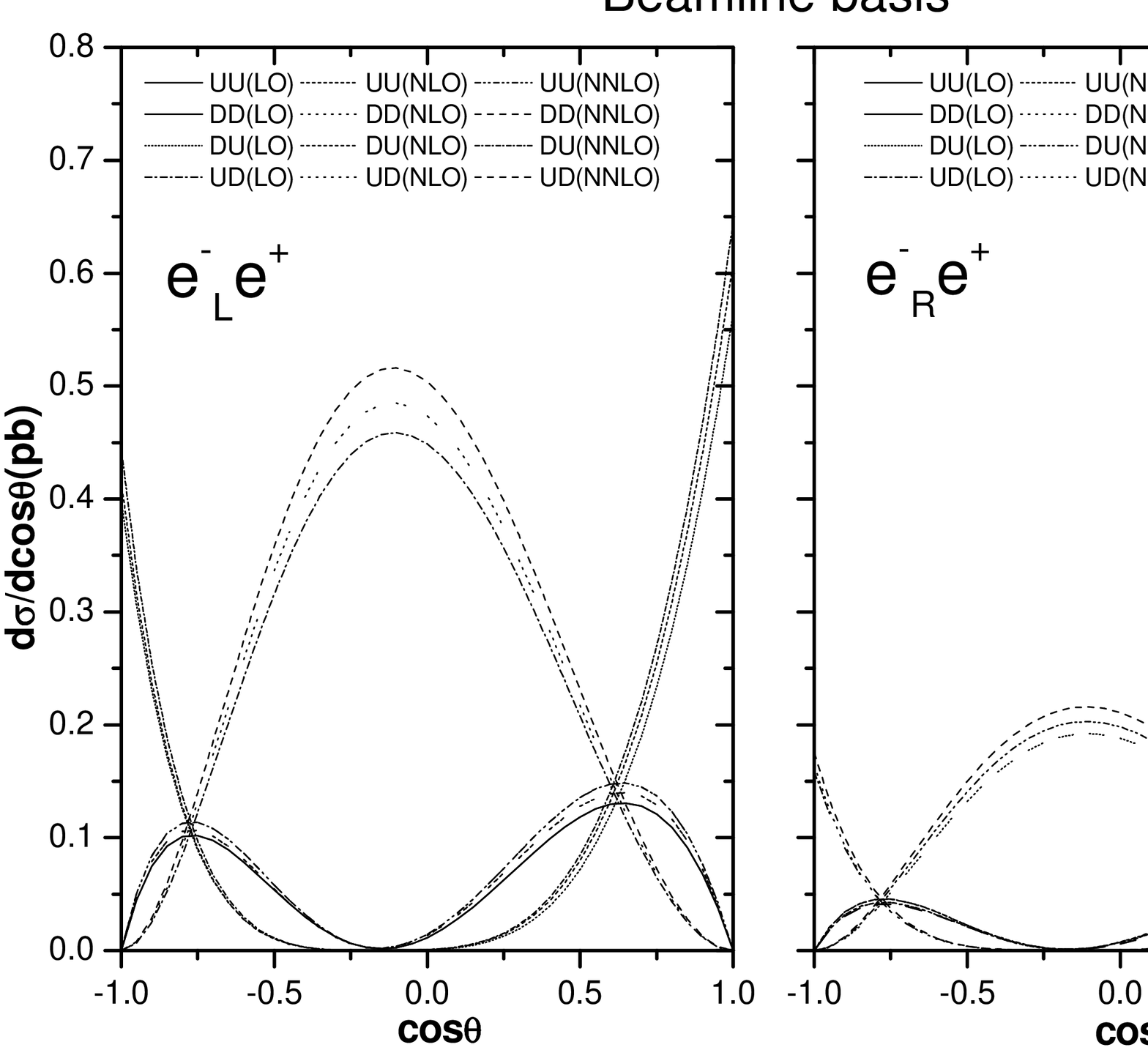, width=400pt}}
\vspace{-3.4cm}\caption[]{The differential cross sections in the
beamline basis for the $e^-_{L,R} e^+ \to t \bar t $ processes
with  the $\uu$(UU), $ \dd$(DD),  $\ud$(UD) and $\du$(DU)
productions, assuming $E_{NR}=\sqrt{s} - 2 m_t  = 5$ GeV.}
\end{figure}

\begin{figure} \label{pt}
\centerline{\epsfig{file=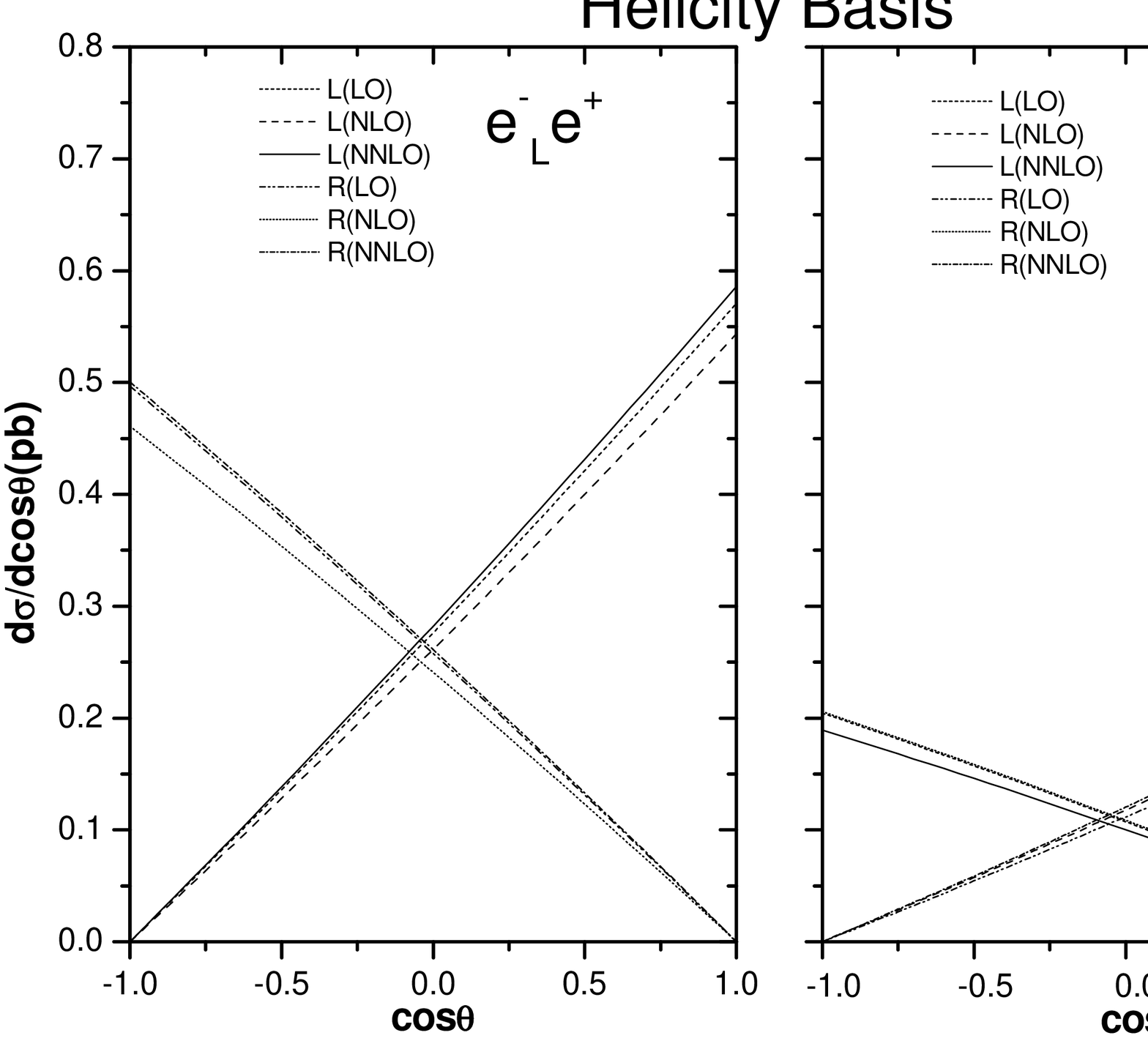, width=400pt}}
\vspace{-3.0cm}\caption[]{The differential cross sections in the
helicity basis for the $e^-_L e^+ \to t_{L,R} \bar t $ and $e^-_R
e^+ \to t_{L,R} \bar t $ processes at $E_{NR}=\sqrt{s} - 2 m_t  =
-1$ GeV.}
\end{figure}

\newpage


\begin{thebibliography}{99}
\bibitem{dec}
I. Bigi, H. Krasemann, Z. Phys. C7, 127(1981); J. K\"uhn, Acta.
Phys. Austr. (Suppl.) XXIV, 203 (1982); I. Bigi, Y. Dokshitzer, V.
Khoze, J. K\"uhn, and P. Zerwas, Phys. Lett. 181B, 157(1986).

\bibitem{tt2}
D. Atwood and A. Soni, Phys. Rev. D45(1992)2405; G.L. Kane, G.A.
Ladinsky and C.P. Yuan, Phys. Rev. D45(1992)124; C. P. Yuan, Phys.
Rev. D45(1992)782; G.A. Ladinsky, Phys. Rev. D46(1992)3789; W.
Bernreuther, O. Nachtmann, P. Overmann and T. Schr\"oder, Nucl.
Phys. B388(1992)53, erratum B406(1993)516; T. Arens and L.M.
Seghal, Nucl. Phys. B393(1993)46; Carl R. Schmidt, PHys. Rev.
D54(1996) 3250; A.Brandenburg, M. Flesch, P. Uwer, Phys. Rev.
D59(1999) 014001; A.Brandenburg, M. Flesch, P. Uwer,
hep-ph/9911249.



\bibitem{imp3}
S. Parke and Y. Shadmi, Phys. Lett. B 387(1996)199.

\bibitem{spi1}
H.X. Liu, C.S. Li, Z.J. Xiao, Phys. Lett. B458(1999)393.

\bibitem{spi}
Michihiro Hori, Yuichiro Kiyo, Jiro Kodaira, Takashi Nasuno,
Stephen Parke, hep-ph/9801370; J. Kodaira, T. Nasuno, S. Parke,
Phys. Rev. D 59(1999)014023; Y. Kiyo, J. Kodaira, K. Morii, T.
Nasuno, S. Parke, Nucl.Phys.Proc.Suppl. 89 (2000) 37.

\bibitem{thr1}
M.J. Strassler, M.E. Peskin, Phys. Rev. D43(1991)1500;  V.S. Fadin
and V.A. Khoze, Yad. fiz. 48(1988) 487; JEPT Lett. 46(1987)525.

\bibitem{imp2}
R. Harlander, M. Je$\dot{\rm z}$abek, J. H. K$\ddot{\rm u}$hn, M.
Peter, Z. Phys. C73 (1997)477.

\bibitem{thr}
M. Je$\dot{\rm z}$abek, J.H. k$\ddot{u}$n, T. Teubner, Z. Phys.
C56(1992) 653;  B.A. Kniehl, A. Sirlin, DESY 92-102; Y. Sumino, K.
Fujii, K. Hagiwara, H. Murayama, C-K. Ng, Phys. Rev. D47(1993) 56;
M. Je$\dot{\rm z}$abek, T. Teubner, Z. Phys. C59(1993) 669; H.
Murayama, Y. Sumino, Phys. Rev. D47(1993) 82.

\bibitem{thr2}
A.H. Hoang, T. Teubner, Phys. Rev. D58 (1998) 114023; A.H. Hoang,
T. Teubner, Phys. Rev. D60 (1999)114027; T. Nagabno, $\Lambda$.
Ota, Y.Sumino, Phys. Rev. D60 (1999)114014; A.H. Hoang {\it et.
al.}, CERN-TH/99-415; M.van Iersel, C.F.M van der Burgh, B.L.G.
Bakker, hep-ph/0010243; M. Beneke, CERN-TH/99-281; S. Su, M. B.
Wise, Phys. Lett. B510(2001)205; A.A. Penin, A.A. Pivovarov, Phys.
Atom. Nucl. 64(2001)275; Yad. Fiz. 64(2001)323; L.W. Stewart, AIP
Conf.Proc. 618 (2002) 395.

\bibitem{edms}
M. Je$\dot{\rm z}$abek, T. Nagano, Y. Sumino, Phys. Rev. D62
(2000)014034.

 \bibitem{pot}
W. Fisher, Nucl. Phys. B129(1977)157; A. Billoire, Phys. Lett.
B92(1980)343; S.N. Gupta and S. Radford, Phys. Rev. D24(1981)2309;
Phy. Rev. D25(1982)3430 (erratum); S.N. Gupta, S.F. Radford and
W.W. Repko, Phys. Rev. D26(1982)3305; M. Peter, Phys. Rev. Lett.
78(1997)602; Nucl. Phys B501(1997)471; Y. Schr$\ddot{\rm o}$der,
Phys. Lett. B447(1999)321.


\bibitem{cpv}
D.Atwood, S. Bar-Shalom, G. Eilam, A. soni Phys. Rept. 347(2001)1;
S. Khalil, IPPP/02/78, DCPT/02/156; S.D. Rindani, hep-ph/0202045.

\bibitem{par}
Particle Data Group, K. Hagiwara et al., Phys. Rev. D 66,
(2002)010001.



\end{thebibliography}
\end{document}